\documentclass[A4paper,unsortedaddress,superscriptaddress,prl,twocolumn,showpacs]{revtex4-1}
\usepackage[pdftex]{graphicx}
\usepackage{graphicx}
\usepackage[latin1]{inputenc}
\usepackage{amsmath}
\usepackage[T1]{fontenc}
\usepackage{color}

\begin{document}
\DeclareGraphicsExtensions{.pdf}

\title{Mapping Image Potential States on Graphene Quantum Dots}

\author{Fabian Craes} 
\email{craes@ph2.uni-koeln.de}
\affiliation{II. Physikalisches Institut, Universit\"{a}t zu K\"{o}ln, Z\"{u}lpicher Stra\ss e 77, 50937 K\"{o}ln, Germany}
\author{Sven Runte}
\affiliation{II. Physikalisches Institut, Universit\"{a}t zu K\"{o}ln, Z\"{u}lpicher Stra\ss e 77, 50937 K\"{o}ln, Germany}
\author{J\"{u}rgen Klinkhammer}
\affiliation{II. Physikalisches Institut, Universit\"{a}t zu K\"{o}ln, Z\"{u}lpicher Stra\ss e 77, 50937 K\"{o}ln, Germany}
\author{Marko Kralj}
\affiliation{Institut za fiziku, Bijeni\v{c}ka 46, 10000 Zagreb, Croatia}
\author{Thomas Michely}
\affiliation{II. Physikalisches Institut, Universit\"{a}t zu K\"{o}ln, Z\"{u}lpicher Stra\ss e 77, 50937 K\"{o}ln, Germany}
\author{Carsten Busse}
\affiliation{II. Physikalisches Institut, Universit\"{a}t zu K\"{o}ln, Z\"{u}lpicher Stra\ss e 77, 50937 K\"{o}ln, Germany}

\date{\today}

\begin{abstract}
Free electron like image potential states are observed in scanning tunneling spectroscopy on graphene quantum dots on Ir(111) acting as potential wells. The spectrum strongly depends on the size of the nanostructure as well as on the spatial position on top, indicating lateral confinement. Analysis of the substructure of the first state by spatial mapping of constant energy local density of states reveals characteristic patterns of confined states. The most pronounced state is not the ground state, but an excited state with a favorable combination of local density of states and parallel momentum transfer in the tunneling process. Chemical gating tunes the confining potential by changing the local workfunction. Our experimental determination of this workfunction allows to deduce the associated shift of the Dirac point.

\end{abstract}


\maketitle

Confinement of electrons in nanostructures leads to quantum size effects as a size-dependent electronic structure and atom-like states (characterized by a set of quantum numbers). Recently, first experiments regarding the confinement of image potential states (IPSs) using the spatial resolution of the scanning tunneling microscope (STM) have been performed \cite{Wahl2003,Borisov2007a, Schouteden2009, Schouteden2010, Schouteden2012, Stepanow2011}, transcending pioneering studies based on two photon photoemission (2PPE) \cite{Fischer1993,Fischer1993a}. IPSs are unoccupied states in an attractive image charge Coulomb potential between the Fermi level $E_{\rm F}$ and the vacuum level $E_{\rm F} + \Phi$ given by the local work function $\Phi$. Perpendicular to the surface they feature a hydrogen-like spectrum (characterized by a quantum number $n$) which converges to $E_{\rm F} + \Phi$ \cite{Echenique1978a}, parallel to it a two dimensional electron gas (2DEG) forms with a continuous distribution of parallel momentum $k$ for the case of extended systems. The resulting states can be labeled $\Psi^{(n)}(k)$ with energies $E^{(n)}(k)$. In STM, IPSs appear as peaks in the local density of states (LDOS) measured while retracting the tip from the surface. As they are Stark-shifted due to the electric field between tip and sample \cite{Crampin2005a} they are often referred to as field emission resonances (FERs). 

Confinement effects for IPSs can be induced by nanostructures fulfilling four conditions: (i) the corresponding potential well must have a sufficient depth of $\Delta \Phi = \Phi_{\rm out}-\Phi_{\rm in}$ \cite{Borisov2007a}, a large $\Delta \Phi$ provides strong confinement; (ii) a well-defined shape; (iii) an established preparation that allows to adjust the size in a wide range; (iv) stability under the high STM bias voltage $U$. Whereas previous work provides fascinating first insights into quantum size effects for IPSs, no study yet matches all four conditions: In \cite{Schouteden2009} a first hint at a size dependence of the energies of IPSs confined to Co islands on Au(111) is visible, however here the size variation was less than an order of magnitude. Atom-like patterns have been observed above stacking-fault tetrahedra on Ag(111) \cite{Schouteden2012}, still $\Delta \Phi$ is so small that the resulting weak confinement only acts on the IPSs lowest in energy. The system NaCl on metal is promising as it shows a large $\Delta \Phi$. However, up to now there is no established method to tune the size of islands with a well-defined shape over a wide range  \cite{Ploigt2007,Schouteden2012c}. In consequence, in these experiments electron confinement has not been observed yet. Strong confinement is found for islands of alkali metals on Cu(100) \cite{Stepanow2011}. In this case, the atomic structure of the islands is unclear, the size cannot be varied, and the clusters are not entirely stable during the measurement. An intriguing feature is the coupling between the IPSs on neighboring nanostructures to molecule-like states \cite{Schouteden2009,Stepanow2011,Feng2008}.

IPSs on graphene (gr) are of special interest: On fundamental grounds, they share a common origin with specific states of related materials \cite{Silkin2009} as the interlayer state of graphite, or superatomic states of fullerenes \cite{Feng2008}. As a consequence of graphene's 2D-character a splitting of the IPSs into $\Psi^{(n^+)}$ and $\Psi^{(n^-)}$ has been predicted for free-standing \cite{Silkin2009} and observed for epitaxial graphene weakly coupled to SiC \cite{Bose2010, remark}. However, for the more strongly interacting gr/Ru(0001) this specific splitting was not observed as the presence of the substrate destroys the 2D-character \cite{Borca2010,Armbrust2012}. Still, in this system the energy of the lowest IPS splits due to the strong corrugation of the carbon sheet which allows a large probability density also between graphene and the substrate. In the system under investigation here, 2PPE could demonstrate parabolic IPSs in the large band gap of the Ir substrate \cite{Niesner2012}. Neither the energetic splitting due to the 2D-character nor due to corrugation were observed.


Here, we demonstrate that confinement of IPSs can be observed in graphene quantum dots (GQDs) on Ir(111). Furthermore, both the width and the depth of the confining potential well can be tuned. The GQDs fulfill all conditions outlined above: (i) A large $\Delta \Phi = \Phi_{\rm{Ir}} - \Phi_{\rm{gr}} = (5.79 \pm 0.10)~{\rm eV} - (4.65 \pm 0.10)~{\rm eV} = (1.1 \pm 0.1)$\,eV \cite{Niesner2012}. Beyond that, $\Delta \Phi$ can be tuned: The intercalation of electron acceptors (as, e.g. O \cite{Granas2012}) between the carbon sheet and its substrate leads to a depletion of charge density in graphene's $\pi$-system, which in turn shifts the Dirac point $E_{\rm D}$ to higher energies \cite{Larciprete2012}, and vice versa for donors. Assuming that the band structure of graphene is rigidly pinned to $E_{\rm F}+\Phi$ \cite{Giovanetti2008}, the work function of intercalated graphene is given by $\Phi_{\rm gr/x} = E_{\rm D} + \Phi_{\rm gr}$, allowing us to change the depth of the potential well by doping. (ii) The GQDs have a well defined polygonal shape which can be determined with atomic precision \cite{Subramaniam2012}. (iii) The GQDs have a size tunable from less than 10~nm$^2$ to electronically equivalent to infinite \cite{Coraux2009}. (iv) The system is stable also for high $U$ due to the strong C-C bonds as well as the good conductivity. Fullerenes \cite{Feng2008} and carbon nanotubes \cite{Schouteden2009} can be viewed as extreme cases of confinement. However, for such systems a tuning of the size over orders of magnitude is impossible and the curved geometry gives a new character to the now strongly hybridized states.

\begin{figure}[h!!]
	\begin{center}
	\includegraphics[width = 1 \linewidth]{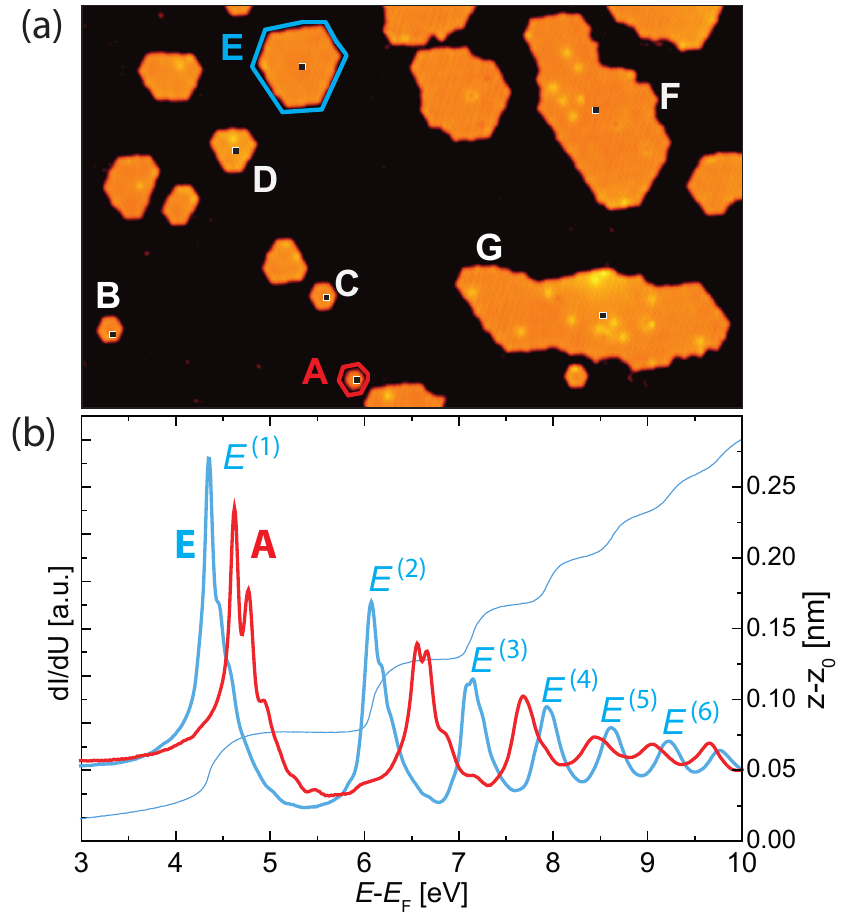}
	\caption{(color online). (a)\,Set of GQDs on Ir(111) [some labeled (A)-(G) with increasing size]; image size 55\,nm$\times$90\,nm, $U=1$V, $I=0.2$nA. (b)\,d$I$/d$U$ spectra on flakes (A) and (E); $U_{\rm{stab}}=1\,$V, $I_{\rm{stab}}=0.2\,$nA; thin (blue) line: simultaneously recorded $z(U)$ on (E).}
		\label{fig:1}
		\end{center}
\end{figure}

Ir(111) substrates are cleaned in ultra high vacuum by cycles of $1.5\,$keV Ar$^{+}$ ion bombardment at 300~K, oxygen firing at $1120\,$K, and annealing at $1470\,$K. A coverage of $\approx 22$~\% of a monolayer of graphene is prepared by decomposition of ethylene in a temperature programmed growth process at 1270\,K \cite{Coraux2009}. The STM tip (made from W) is virtually grounded and the sample is put to $U$ leading to a tunneling current $I$. Energies are given by $E-E_{\rm{F}}= e \cdot U$. IPSs are investigated by measuring the differential conductivity in form of both d$I$/d$U$($E-E_{\rm F}$) point spectra and constant energy maps in constant current mode (stabilization values $U_{\rm stab}$, $I_{\rm stab}$) with active feedback loop by using the lock-in technique ($f=1.317\,$kHz, $U_{\rm{mod}}=14\,$mV) which together with the sample temperature of 5.3~K leads to an energy resolution of $\delta E = 25$~meV \cite{Morgenstern2003}. Active feedback is important as moving the tip in $z$ direction during point spectroscopy compensates for the Stark shift within one measurement \cite{Foerster2011}. The tip traces $z(U)$ were recorded parallel to the spectra [see Fig.\,\ref{fig:1}\,(b)]. All data is taken in a background pressure lower than $1\times10^{-11}$\,mbar. Data analysis is performed using WSxM \cite{Horcas2007}.

Fig.\,\ref{fig:1}\,(a) shows GQDs with sizes from less than 10~nm$^2$ to several hundred nm$^2$, see also \cite{Coraux2009}. The d$I$/d$U$-spectra taken with the same microscopical tip \cite{Suga1999} at the center of the flakes labeled in Fig.\,\ref{fig:1}\,(a) show very pronounced  features [Fig.\,\ref{fig:1}\,(b)]. A set of spectra taken along the diameter of a selected island is presented in the supplementary material \cite{supp}. We attribute the discrete peaks to the energies $E^{(n)}$ of the sequence of IPSs with order $n=1,2,...$. These energies are down-shifted with respect to pristine Ir(111) due to the significantly smaller $\Phi$. The spectra on the differently sized flakes are also shifted with respect to each other by as much as $\Delta E \approx 0.67\,$~eV between the largest and the smallest flake for both $E^{(2)}$ and $E^{(3)}$. Such a large difference cannot be due to a variation of the Stark shift \cite{supp}. In addition, individual peaks show a clear substructure, especially for the smaller island (red line). This is in contrast to the smooth peaks observed for extended systems \cite{supp}. 

\begin{figure}
	\begin{center}
	\includegraphics[width = 1 \linewidth]{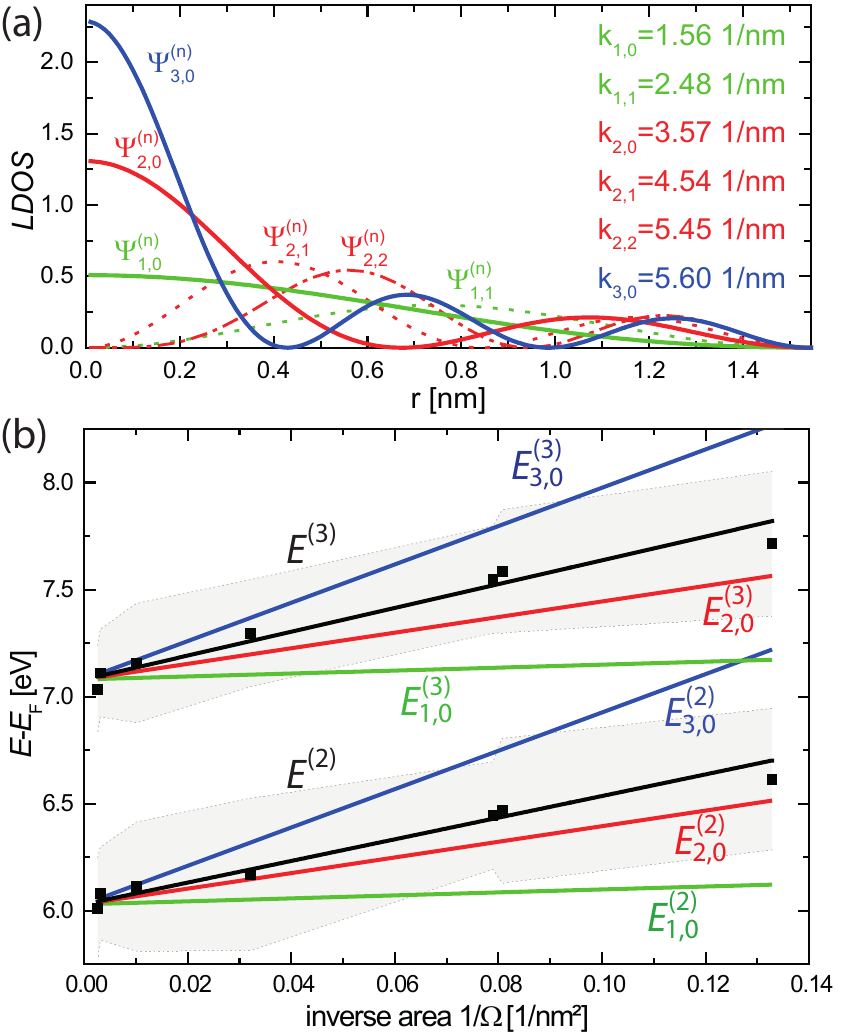}
	\caption{(color online). (a)\,Normalized LDOS and $k_{m,l}$ for a model island with $\Omega=7.5$~nm$^2$, i.e. $r=1.545\,$nm (right border of the plot). (b)\,Energies $E^{(2)}$ and $E^{(3)}$ depending on inverse area, measured on (A)-(G) in Fig.\,\ref{fig:1}\,(a) with parameters from Fig.\,\ref{fig:1}\,(b); peak FWHM (grey shading) and linear fits (black lines); calculated energies $E^{(n)}_{m,l}$, see labels.}
		\label{fig:2} 
		\end{center}
\end{figure}

A dependence of the energy of electronic states on the size of the system as well as the occurrence of discrete eigenstates evidenced by the peak substructure is a clear fingerprint of lateral quantum confinement. To treat this analytically, we approximate our hexagonal GQDs by an infinite cylindrical potential well with radius $r$ \cite{Barke2006}. The eigenfunctions in polar coordinates $(\rho, \phi)$ are then $\Psi^{(n)}_{m,l} \propto J_l(k_{m,l} \rho) e^{\pm i l \phi}$, where $J_l$ is the spherical Bessel function of the first kind with order $l$ \cite{Crommie1993}. The continuous distribution of $k$ breaks down into discrete values $k_{m,l}$. Due to the confinement, $\Psi^{(n)}_{m,l}$ must have a node for $\rho=r$, leading to the condition $k_{m,l} r =z_{m,l}$ with $z_{m,l}$ the $m$-th zero of $J_l$, i.e., the eigenstates can be characterized by two additional quantum numbers $m$ and $l$, hence $\Psi^{(n)}_{m,l}$. Their energies are given by $E^{(n)}_{m,l}=E^{(n)}_0+(\hbar^2\pi/2)\cdot z_{m,l}^2/(\Omega \cdot m^*)$, with $E^{(n)}_0$ the energy of the state on extended graphene, $\Omega$ the flake area, and $m^*$ the effective electron mass. For $r \rightarrow \infty$ this equation converges to the dispersion relation for free electrons. Radial cuts through the normalized probability density $\Psi^{(n)*}_{m,l}\Psi^{(n)}_{m,l}$ of the first six eigenstates are shown in Fig.\,\ref{fig:2}\,(a). This plot can be used to explain the substructure of the peaks in Fig.\,\ref{fig:1}\,(a): A spectrum taken at a point $r$ will pick up LDOS from several states at the respective $E^{(n)}_{m,l}$. Note that even though we took all spectra in the center of the islands ($r=0$), especially for the case of small dots a small contribution of states with $l \neq 0$ has to be expected since the spatial resolution decreases when the tip-sample distance $z_0$ is no longer small with respect to $r$.

We fitted the spectra phenomenologically by a sequence of $n$ Voigt functions. Fig.\,\ref{fig:2}\,(b) shows the position of the peaks $E^{(n)}$ for $n=2,3$ (black squares) and their full width at half maximum (grey shading) against $\Omega^{-1}$ for the whole set of flakes shown in Fig.\,\ref{fig:1}\,(a). The state $n=1$ is disregarded as it is still strongly influenced by the density of states of the substrate \cite{Lin2007a}. We compare a linear fit to the data (black lines) with the expected behavior for $E^{(n)}_{m,l}$ ($m^*\approx  m_e$ \cite{Niesner2012}, $E_0^{(n)}$ from fit), focusing on states with $l=0$ [labeled lines in Fig.\,\ref{fig:2}\,(b)] as we have measured the spectra at $r=0$ [see also Fig.\,\ref{fig:1}\,(a)]. Obviously the data fits best to $\Psi^{(n)}_{2,0}$, both for $n=2$ and $n=3$. This is surprising as this is not the ground state. 

The dominance of $\Psi^{(n)}_{2,0}$ cannot be explained by the LDOS at the center of the flake, as this quantity increases with $m$ for $l=0$ [see Fig.\,\ref{fig:2}\,(a)].  For a correct interpretation, one has to take into account that tunneling is most probable for electrons with vanishing $k$. However, for the electrons confined above the GQDs, $k_{\rm m,l}$ increases with $m$ [see Fig.\,\ref{fig:2}\,(a)], making them less accessible for STS. In consequence, whereas the LDOS in the center increases with the order of the state, the contribution to the tunneling density of states (TDOS, e.g. \cite{Wehling2008}) measured in STM decreases: {$\rm{TDOS}=\rm{LDOS}\cdot \exp(-z_0/\lambda)$} \cite{Zhang2008}, with $\lambda^{-1}=2\sqrt{2m_e\Phi/\hbar^2+k_{m,l}^2}$ and $k_{m,l}=z_{m,l}/r$. Evaluating this formula for the model island with $\Omega=7.5\,$nm$^2$ indeed yields that $\Psi^{(n)}_{2,0}$ dominates for $0.36~{\rm nm} < z_0 < 1.04$~nm. It is reasonable to assume that $z_0$ in our experiment is within these boundaries. The considerations on the tunneling probability and thus preferred states for tunneling may also connect to previous publications like \cite{Li1998a}, where the decrease in peak intensity could also be explained by less probable tunneling due to increased $k$ instead of an {\it ad hoc} assumption of a peak broadening increasing with energy.

\begin{figure}[htbp]
	\begin{center}
	\includegraphics[width = 0.9 \linewidth]{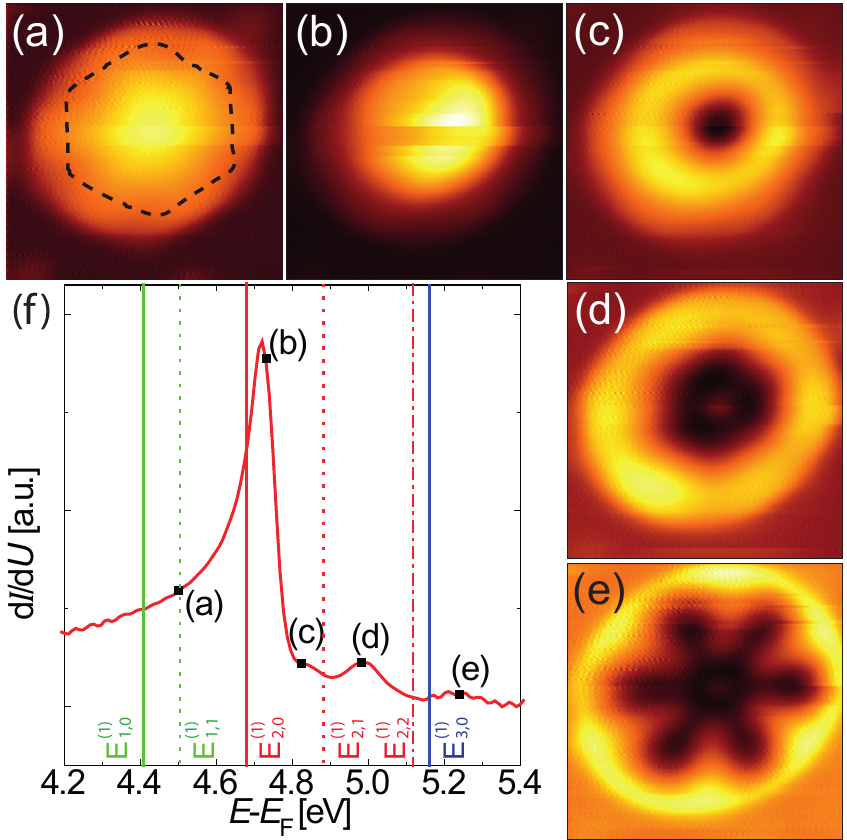}
	\caption{(color online). (a)-(e)\,IPS $n=1$ by 2D constant energy mapping of d$I$/d$U$ on a $\Omega=11\,$nm$^2$ sized GQD. (a)-(e): $E-E_{\rm{F}}=4.50\,$eV, $4.73\,$eV, $4.82\,$eV, $4.98\,$eV, $5.22\,$eV; $I=0.2\,$nA; image size: all $5.7\,\times 5.7\,$nm$^2$; dashed line in (a): topography contour at $E-E_{\rm{F}}=0.2\,$eV. (f)\,d$I$/d$U$($E-E_F$) spectrum of $n=1$ with energies of maps (a)-(e) (black squares), $E^{(1)}_{m,l}$ indicated by solid (dashed) lines for $l=0$ ($l\neq0$); $E^{(1)}_{0}$ as $E^{(2,3)}_{0}$ in Fig.\,\ref{fig:2}\,(b).}
		\label{fig:3}
		\end{center}
\end{figure}

The spatial modulation of the LDOS can be resolved by d$I$/d$U$ mapping on a hexagonal flake for $n=1$, see Fig.\,\ref{fig:3}. For higher $n$, our resolution in space and energy was not sufficient to detect significant spatial variation of the LDOS, similar to \cite{Schouteden2012}. The d$I/$d$U$-spectrum [Fig.\,\ref{fig:3}\,(f)] shows a substructure equivalent to the one in Fig.\,\ref{fig:2} (b), which does not have exactly the same shape due to different experimental parameters (including tip shape). Again, the maximum corresponds to $E^{(1)}_{2,0}$. The maps at selected energies can again be understood on the basis of Fig.\,\ref{fig:2} (a): The state shown in Fig.~\ref{fig:3}(a) shows a broad maximum resembling $\Psi^{(n)}_{1,0}$, (b) is more peaked in the center like $\Psi^{(n)}_{2,0}$, and (c) and (d) have vanishing intensity in the center like $\Psi^{(n)}_{2,1}$. Strictly speaking, however, the patterns we observe are not pure states, but a superposition of several neighboring states. Note that in Fig.~\ref{fig:3} (d) and (e) the breaking of the cylindrical symmetry by the hexagonal shape of the flake becomes evident. Similar patterns have been observed on GQDs as a result of confinement of low energy occupied states \cite{Subramaniam2012, Phark2011, Hamalainen2011, Altenburg2012}. Finally, the inversion of contrast in Fig.\,\ref{fig:3}\,(e) indicates that the LDOS at this energy is dominated by Ir IPSs, drawing our attention to the interaction of the IPS-2DEGs above graphene and Ir(111). Mapping along lines cutting through a GQD (see \cite{supp}) shows that the energy shift for $n=1$ across the boundary region between $\Phi_{\rm Ir}$ and $\Phi_{\rm gr}$ is rather abrupt, whereas all higher orders show a more continuous change. This indicates a suppressed interaction for $n=1$ and interacting IPS-2DEGs of the flake and the substrate for $n>1$.

\begin{figure}
	\begin{center}
	\includegraphics[width = 1 \linewidth]{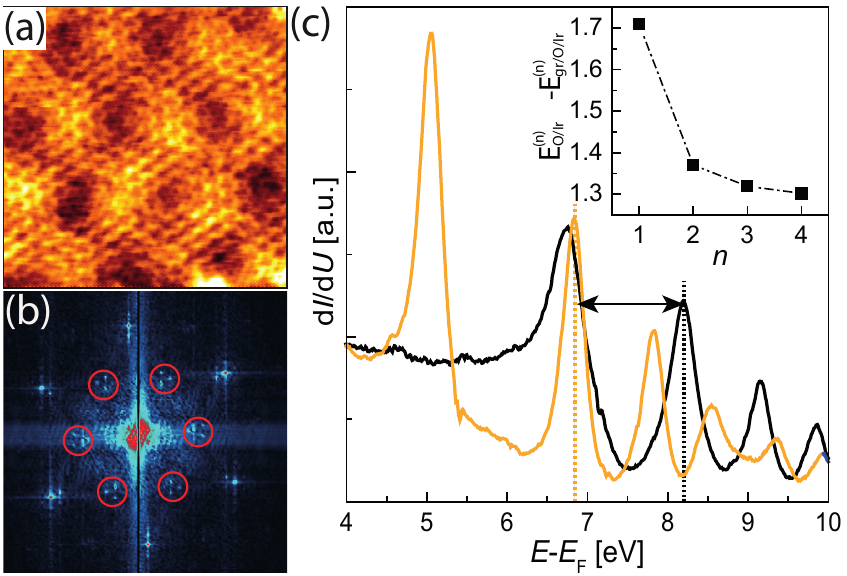}
	\caption{(color online). (a)\,topography: O intercalated GQD, size $7.6\times7.6\,$nm$^2$, U=110mV, I=0.08\,nA; (b)\,FT-STM of gr/O/Ir(111), circles: $(\sqrt{3}\times\sqrt{3})$R$30^{\circ}$-O intercalation superstructure. (c)\,IPS spectra on gr/O/Ir(111) \,(yellow) and on O/Ir(111)\,(black), $\Delta \Phi$ indicated by black arrow; inset: $\Delta E^{(n)}=E_{\rm{O/Ir}}^{(n)} - E_{\rm{gr/O/Ir}}^{(n)}$.}
		\label{fig:4} 
		\end{center}
\end{figure}

In the following, we will exemplify chemical gating of graphene for the case of O. The sample preparation is extended by exposure to $750\,$L of O$_2$ at $430\,$K, leading to O intercalation for all but the smallest flakes \cite{Granas2012}. In order to demonstrate the effect of intercalation most clearly, we will focus on the largest flakes which are representative for extended graphene. In the respective STM image [Fig.~\ref{fig:4} (a)], three superimposed structures can be made out: The graphene honeycomb structure is faintly visible inside the dark depressions. The more pronounced small-scale pattern is a $(\sqrt{3} \times \sqrt{3})$R$30^{\circ}$ structure with respect to Ir(111) as evidenced in the the corresponding Fourier transform (FT) shown in Fig.~\ref{fig:4} (b), see the circled spots. It is formed by the adsorption of the intercalated O to Ir(111), indicating a coverage of 0.33~ML. The large scale pattern (satellite spots in the FT) is due to the moiré structure formed by the incommensurate lattices of graphene and Ir(111). In Fig.\,\ref{fig:4} (c) we compare spectra on gr/O/Ir (yellow) and on O/Ir(111) (black). We derive $\Delta\Phi=(1.3 \pm 0.1)\,$eV from a plot $\Delta E^{(n)}=E_{\rm{O/Ir}}^{(n)} - E_{\rm{gr/O/Ir}}^{(n)}$ versus $n$ \cite{Lin2007a}, see inset of Fig.\,\ref{fig:4} (c). Note that especially for $n=1$ a large deviation form this value results which is due to the interaction of the lowest IPS with the substrate \cite{Lin2007a}, as already mentioned above. We obtain $\Phi_{\rm{gr/O/Ir}}=( 5.1 \pm 0.1)\,$eV, which has to be compared with $\Phi_{\rm{gr/Ir}}=(4.7\pm0.1)$\,eV \cite{Ivanov1976,Niesner2012}. 

According to \cite{Giovanetti2008}, we deduce $\Delta E_{\rm D}= E_{\rm D, gr/O/Ir} - E_{\rm D, gr/Ir} = \Phi_{\rm gr/O/Ir} - \Phi_{\rm gr/Ir}= (0.4 \pm 0.1)$~eV. This nicely agrees with  $\Delta E_{\rm D}=0.3$~eV implied by a recent photoemission study \cite{Larciprete2012} (linearly interpolated using $\Delta E_{\rm D}=0.6\,$eV for 0.6\,ML). In consequence, our determination of the local work function provides direct access $E_{\rm D}$, which is often hard to determine by other methods. As an example, for gr/Ir(111) the LDOS determined from STS does not show a pronounced dip at $E_{\rm D}$ \cite{Li2013}. A determination of $E_{\rm D}$ via $\Phi$ can be especially useful for mapping the doping level in inhomogeneous graphene systems.

As an outlook, the suitability of graphene as building material for nanostructures opens new possibilities for the investigation of laterally confined IPSs, as there are various methods on how to prepare rationally designed architectures: Graphene can be cut by STM lithography \cite{Tapaszto2008}, allowing complicated well geometries. A clever choice of the hydrocarbon precursor leads to the formation of superperiodic structures \cite{Bieri2009} which may give rise to backfolding of the IPS bands. One can envision that it is possible to move GQDs with an STM tip, allowing exact studies of the interaction of neighboring quantum wells. Such structures would resemble a diatomic molecule, where the participating quantum wells can even have different energy levels due to a variation in size or in doping. 

Summing up our results, the large difference in workfunction between graphene and Ir(111), the well defined shape of the nanostructures as well as their large size variation and high stability enabled us to demonstrate confinement effects of IPSs. We have shown that the energy spectrum depends on the size of the GQDs and evolves into a series of atom-like states, which is dominated by a state other than the ground state due to an interplay of density of states and parallel momentum transfer in the tunneling process. Intercalating extended graphene with an electron acceptor introduces an additional degree of freedom as this allows tuning of $\Phi$. The determination of the local workfunction allows to deduce the local doping level $E_{\rm D}$. Our findings open new possibilities for the study of quantum size effects as graphene is a very flexible building material for nanostructures. 

This work is supported by DFG through SFB 608, the projects Bu2197/2-1, INST 2156/514-1, and the BCGS, the EU through the project GRENADA, and DAAD-MZOS via the project ``Electrons in two dimensions''.


\begin{thebibliography}{10}

\bibitem{Wahl2003}
P. Wahl, M.~A. Schneider, L. Diekh\"oner, R. Vogelgesang, and K. Kern, Phys.
  Rev. Lett. {\bf 91},  106802  (2003).

\bibitem{Borisov2007a}
A.~G. Borisov, T. Hakala, M.~J. Puska, V.~M. Silkin, N. Zabala, E.~V. Chulkov,
  and P.~M. Echenique, Phys. Rev. B {\bf 76},  121402  (2007).

\bibitem{Schouteden2009}
K. Schouteden and C. Van~Haesendonck, Phys. Rev. Lett. {\bf 103},  266805
  (2009).

\bibitem{Schouteden2010}
K. Schouteden, A. Volodin, D.~A. Muzychenko, M.~P. Chowdhury, A. Fonseca, J.~B.
  Nagy, and C. Van~Haesendonck, Nanotechnology {\bf 21},  485401  (2010).

\bibitem{Schouteden2012}
K. Schouteden and C. Van~Haesendonck, Phys. Rev. Lett. {\bf 108},  076806
  (2012).

\bibitem{Stepanow2011}
S. Stepanow, A. Mugarza, G. Ceballos, P. Gambardella, I. Aldazabal, A.~G.
  Borisov, and A. Arnau, Phys. Rev. B {\bf 83},  115101  (2011).

\bibitem{Fischer1993}
R. Fischer, S. Schuppler, N. Fischer, T. Fauster, and W. Steinmann, Phys. Rev.
  Lett. {\bf 70},  654  (1993).

\bibitem{Fischer1993a}
R. Fischer, T. Fauster, and W. Steinmann, Phys. Rev. B {\bf 48},  15496
  (1993).

\bibitem{Echenique1978a}
P.~M. Echenique and J.~B. Pendry, J. Phys. C: Solid State Phys. {\bf 11},  2065
   (1978).

\bibitem{Crampin2005a}
S. Crampin, Phys. Rev. Lett. {\bf 95},  046801  (2005).

\bibitem{Ploigt2007}
H.-C. Ploigt, C. Brun, M. Pivetta, F.~m.~c. Patthey, and W.-D. Schneider, Phys.
  Rev. B {\bf 76},  195404  (2007).

\bibitem{Schouteden2012c}
K. Lauwaet, K. Schouteden, E. Janssens, C.~V. Haesendonck, and P. Lievens, J.
  Phys. Condens. Matter {\bf 24},  475507  (2012).

\bibitem{Feng2008}
M. Feng, J. Zhao, and H. Petek, Science {\bf 320},  359  (2008).

\bibitem{Silkin2009}
V.~M. Silkin, J. Zhao, F. Guinea, E.~V. Chulkov, P.~M. Echenique, and H. Petek,
  Phys. Rev. B {\bf 80},  121408  (2009).

\bibitem{Bose2010}
S. Bose, V.~M. Silkin, R. Ohmann, I. Brihuega, L. Vitali, C.~H. Michaelis, P.
  Mallet, J.~Y. Veuillen, M.~A. Schneider, E.~V. Chulkov, P.~M. Echenique, and
  K. Kern, New J. Phys. {\bf 12},  023028  (2010).

\bibitem{remark}
Note that in \cite{Silkin2009} and \cite{Bose2010} a shortened notation is
  used, e.g. $1^+$ for $\Psi^{(1^+)}$.

\bibitem{Borca2010}
B. Borca, S. Barja, M. Garnica, D. S\'anchez-Portal, V.~M. Silkin, E.~V.
  Chulkov, C.~F. Hermanns, J.~J. Hinarejos, A.~L. V\'azquez~de Parga, A. Arnau,
  P.~M. Echenique, and R. Miranda, Phys. Rev. Lett. {\bf 105},  036804  (2010).

\bibitem{Armbrust2012}
N. Armbrust, J. G\"udde, P. Jakob, and U. H\"ofer, Phys. Rev. Lett. {\bf 108},
  056801  (2012).

\bibitem{Niesner2012}
D. Niesner, T. Fauster, J.~I. Dadap, N. Zaki, K.~R. Knox, P.-C. Yeh, R.
  Bhandari, R.~M. Osgood, M. Petrovi\ifmmode~\acute{c}\else \'{c}\fi{}, and M.
  Kralj, Phys. Rev. B {\bf 85},  081402  (2012).

\bibitem{Granas2012}
E. Gr{\aa}n{\"a}s, J. Knudsen, U.~A. Schr\"{o}der, T. Gerber, C. Busse, M.~A.
  Arman, K. Schulte, J.~N. Andersen, and T. Michely, ACS Nano {\bf 6},  9951
  (2012).

\bibitem{Larciprete2012}
R. Larciprete, S. Ulstrup, P. Lacovig, M. Dalmiglio, M. Bianchi, F. Mazzola, L.
  Hornekaer, F. Orlando, A. Baraldi, P. Hofmann, and S. Lizzit, ACS Nano {\bf
  6},  9551  (2012).

\bibitem{Giovanetti2008}
G. Giovannetti, P.~A. Khomyakov, G. Brocks, V.~M. Karpan, J. van~den Brink, and
  P.~J. Kelly, Phys. Rev. Lett. {\bf 101},  026803  (2008).

\bibitem{Subramaniam2012}
D. Subramaniam, F. Libisch, Y. Li, C. Pauly, V. Geringer, R. Reiter, T.
  Mashoff, M. Liebmann, J. Burgd\"orfer, C. Busse, T. Michely, R. Mazzarello,
  M. Pratzer, and M. Morgenstern, Phys. Rev. Lett. {\bf 108},  046801  (2012).

\bibitem{Coraux2009}
J. Coraux, A.~T. N'Diaye, M. Engler, C. Busse, D. Wall, N. Buckanie, F.-J.
  {Meyer zu Heringdorf }, R. van Gastel, B. Poelsema, and T. Michely, New J.
  Phys. {\bf 11},  023006  (2009).

\bibitem{Morgenstern2003}
M. Morgenstern, Surface Review and Letters {\bf 10},  933  (2003).

\bibitem{Foerster2011}
D.~F. F{\"o}rster, Ph.D. thesis, University of Cologne, 2011.

\bibitem{Horcas2007}
I. Horcas, R. Fernandez, J.~M. Gomez-Rodriguez, J. Colchero, J. Gomez-Herrero,
  and A.~M. Baro, Rev. Sci. Instr. {\bf 78},    (2007).

\bibitem{Suga1999}
Y. Suganuma and M. Tomitori, Surf. Sci. {\bf 438},  311   (1999).

\bibitem{supp}
See Supplemental Material at [URL will be inserted by publisher] for a
  discussion of the influence of Stark shift, the IPS-2DEG interaction and the
  missing substructure in d$I$/d$U$ spectra on large GQDs.

\bibitem{Barke2006}
H. H{\"o}vel and I. Barke, Prog. Surf. Sci. {\bf 81},  53   (2006).

\bibitem{Crommie1993}
M. Crommie, C. Lutz, and D. Eigler, Science {\bf 262},  218  (1993).

\bibitem{Lin2007a}
C.~L. Lin, S.~M. Lu, W.~B. Su, H.~T. Shih, B.~F. Wu, Y.~D. Yao, C.~S. Chang,
  and T.~T. Tsong, Phys. Rev. Lett. {\bf 99},  216103  (2007).

\bibitem{Wehling2008}
T.~O. Wehling, I. Grigorenko, A.~I. Lichtenstein, and A.~V. Balatsky, Phys.
  Rev. Lett. {\bf 101},  216803  (2008).

\bibitem{Zhang2008}
Y. Zhang, V.~W. Brar, F. Wang, C. Girit, Y. Yayon, M. Panlasigui, A. Zettl, and
  M.~F. Crommie, Nature Physics {\bf 4},  627  (2008).

\bibitem{Li1998a}
J. Li, W.-D. {Schneider}, R. {Berndt}, and S. {Crampin}, Phys. Rev. Lett. {\bf
  80},  3332  (1998).

\bibitem{Phark2011}
S.-h. Phark, J. Borme, A.~L. Vanegas, M. Corbetta, D. Sander, and J. Kirschner,
  ACS Nano {\bf 5},  8162  (2011).

\bibitem{Hamalainen2011}
S.~K. H\"am\"al\"ainen, Z. Sun, M.~P. Boneschanscher, A. Uppstu, M. Ij\"as, A.
  Harju, D. Vanmaekelbergh, and P. Liljeroth, Phys. Rev. Lett. {\bf 107},
  236803  (2011).

\bibitem{Altenburg2012}
S.~J. Altenburg, J. Kr\"oger, T.~O. Wehling, B. Sachs, A.~I. Lichtenstein, and
  R. Berndt, Phys. Rev. Lett. {\bf 108},  206805  (2012).

\bibitem{Ivanov1976}
V.~P. Ivanov, G.~K. Boreskov, V.~I. Savchenko, W.~F. {Egelhoff Jr.}, and W.~H.
  Weinberg, Surf. Sci. {\bf 61},  207   (1976).

\bibitem{Li2013}
Y. Li, D. Subramaniam, N. Atodiresei, P. Lazi\'{c}, V. Caciuc, C. Pauly, A.
  Georgi, C. Busse, M. Liebmann, S. Bl\"{u}gel, M. Pratzer, M. Morgenstern, and
  R. Mazzarello, Adv. Mater. {\bf 25},  1967  (2013).

\bibitem{Tapaszto2008}
L. Tapaszto, G. Dobrik, P. Lambin, and L.~P. Biro, Nat. Nanotechnol. {\bf 3},
  397  (2008).

\bibitem{Bieri2009}
M. Bieri, M. Treier, J. Cai, K. Ait-Mansour, P. Ruffieux, O. Groning, P.
  Groning, M. Kastler, R. Rieger, X. Feng, K. M\"ullen, and R. Fasel, Chem.
  Commun.  6919  (2009).

\end{thebibliography}
\end{document}